\documentclass[aps,prl,amsmath,twocolumn,superscriptaddress,letterpaper,floatfix]{revtex4-1}
\usepackage{graphicx,color}
\usepackage{verbatim}
\usepackage{amssymb}   
\usepackage{amsmath}
\usepackage{amsfonts}
\usepackage{mathdots}
\usepackage{hyperref}
\usepackage{epsfig}

\usepackage{braket}
\usepackage{bm}

\begin{document}

\title{Moir\'{e} Quantum Layer Hall Effect}
  \author{Jin-Xin Hu}
\affiliation{Department of Physics, Hong Kong University of Science and Technology, Clear Water Bay, Hong Kong, China}

 \author{Wen-Bo Dai}
\affiliation{Department of Physics, Hong Kong University of Science and Technology, Clear Water Bay, Hong Kong, China}

\author{X. C. Xie}
	\affiliation{International Center for Quantum Materials, School of Physics, Peking University, Beijing 100871, China}
	\affiliation{Interdisciplinary Center for Theoretical Physics and Information Sciences, Fudan University, Shanghai 200433, China}
  \affiliation{Hefei National Laboratory, Hefei 230088, China}

	\date{\today}
	\begin{abstract}
Even-layer topological antiferromagnetic thin films such as MnBi$_2$Te$_4$ exhibit an unconventional layer Hall effect (LHE) arising from a layer-locked Berry curvature. This effect manifests as a layer-polarized anomalous Hall effect that can be controlled by a vertical displacement field. However, achieving a quantized version of the LHE remains experimentally challenging. In this work, we propose that moir\'{e} engineering, using an electrostatic superlattice potential from control layers, can realize the quantum layer Hall effect (QLHE). Specifically, by placing moir\'{e} control layers in close proximity to the surfaces of a MnBi$_2$Te$_4$ thin film without breaking $PT$ symmetry, the resulting system exhibits a vanishing net Hall conductance but a quantized layer Hall conductance of $e^2/h$. A weak gate electric field then breaks $PT$ symmetry and drives the system into a layer-polarized quantum anomalous Hall (QAH) phase. Furthermore, when a single control layer is introduced on only one surface, the layer-polarized QAH emerges even in the absence of an external electric field. We identify the microscopic origin of the QLHE as an emergent layer-$U(1)$ gauge field generated by the scalar moir\'{e} potential, which in turn produces opposite periodic pseudomagnetic fields on the top and bottom layers. Our work provides a feasible pathway toward realizing the QLHE in antiferromagnetic thin films.

	\end{abstract}
	\pacs{}	
	\maketitle

{\emph {Introduction.}}---The quantum anomalous Hall (QAH) effect, a hallmark of topological phenomena in condensed matter physics, is characterized by a quantized Hall conductance in the absence of an external magnetic field~\cite{haldane1988model,chang2023colloquium,chang2013experimental,yu2010quantized,qiao2010quantum,qiao2014quantum,chang2015high,deng2020quantum,zhao2020tuning,liu2008quantum,weng2015quantum}. It was first experimentally realized in magnetically doped topological insulator thin films, where ferromagnetic order provides the necessary net magnetization to break time-reversal symmetry~\cite{chang2013experimental,chang2015high}. More recently, moir\'{e} materials such as twisted bilayer graphene~\cite{sharpe2019emergent,serlin2020intrinsic,liu2021theories,zhang2019twisted} and transition metal dichalcogenides~\cite{li2021quantum,xu2023observation,zeng2023thermodynamic,cai2023signatures,park2023observation} have also been shown to host the QAH phase, driven by topological flat bands (with orbital magnetization) and strong electron correlations. This has naturally fostered the prevailing view that a net magnetization is required to realize the QAH phase~\cite{liu2016quantum}. However, this reliance on net magnetization not only poses practical challenges for realizing and manipulating QAH phases~\cite{tokura2019magnetic}, but also raises a fundamental question: can the QAH phase be realized in compensated antiferromagnets (AFMs) that possess no net spin magnetization?

\begin{figure}
		\centering
		\includegraphics[width=1\linewidth]{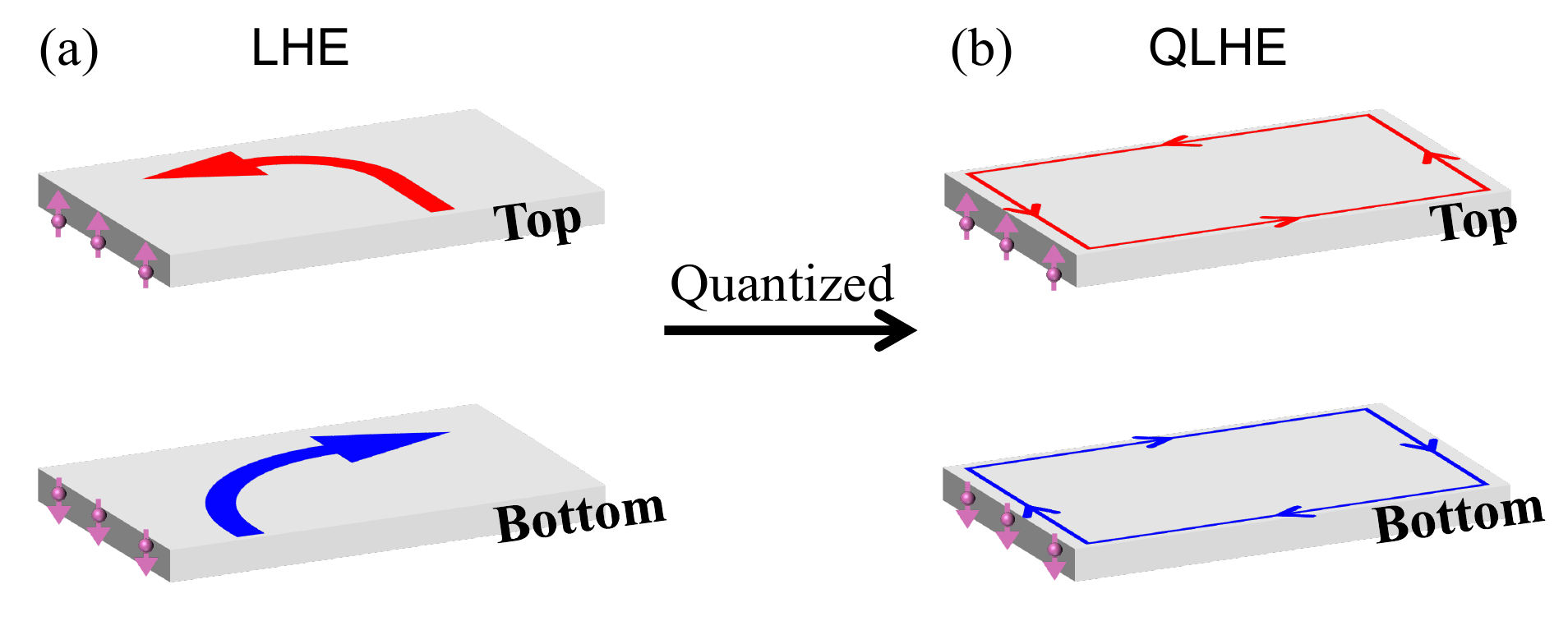}
`		\caption{Conceptual illustration of the layer Hall effect (LHE) and its quantized counterpart, the quantum layer Hall effect (QLHE). The top and bottom layers carry opposite magnetizations, forming a compensated AFM. (a) The LHE is characterized by opposite Hall responses associated with the top and bottom layers. (b) Upon quantization, the opposite layer Hall responses evolve into counter-propagating chiral edge states localized on opposite layers, giving rise to the QLHE.} 
		\label{fig:fig1}
\end{figure}

Nevertheless, a new paradigm has recently emerged from a class of topological antiferromagnetic multilayers such as MnBi$_2$Te$_4$ (MBT). In even-layer MBT films, a distinct phenomenon, the layer Hall effect (LHE), challenges this conventional wisdom~\cite{gao2021layer,peng2023intrinsic,chen2024layer,han2025layer}. In such a system, adjacent layers carry opposite spin polarizations, and the corresponding interlayer N\'{e}el order results in a compensated magnetization~\cite{liu2020robust,li2019intrinsic,qiu2025observation,li2024progress,lian2025antiferromagnetic}. Consequently, electrons from the top and bottom layers are deflected toward opposite edges, generating a layer-contrasting transverse current~\cite{chen2024layer}(see Fig.\ref{fig:fig1}(a)). Remarkably, owing to their atomically thin nature, even-layer MBT films provide an electrically tunable platform in which the LHE manifests itself as an anomalous Hall response~\cite{gao2021layer}. A perpendicular electric field breaks the combined parity-time (\(PT\)) symmetry, lifting the cancellation of Berry curvature between opposite layers and generating a measurable layer-polarized Hall signal.
Despite this advance, it is important to note that the observed Hall response is not yet quantized to $e^2/h$; a genuine QAH state in such collinear AFMs has remained elusive. Although recent theoretical proposals have suggested promising routes toward realizing QAH in compensated AFMs through disorder~\cite{dai2022quantum} or magnetic pinning~\cite{liang2025chern}, they require additional disorder or magnetic-order engineering, posing challenges for experimental implementation.

\begin{figure}
		\centering
		\includegraphics[width=1.0\linewidth]{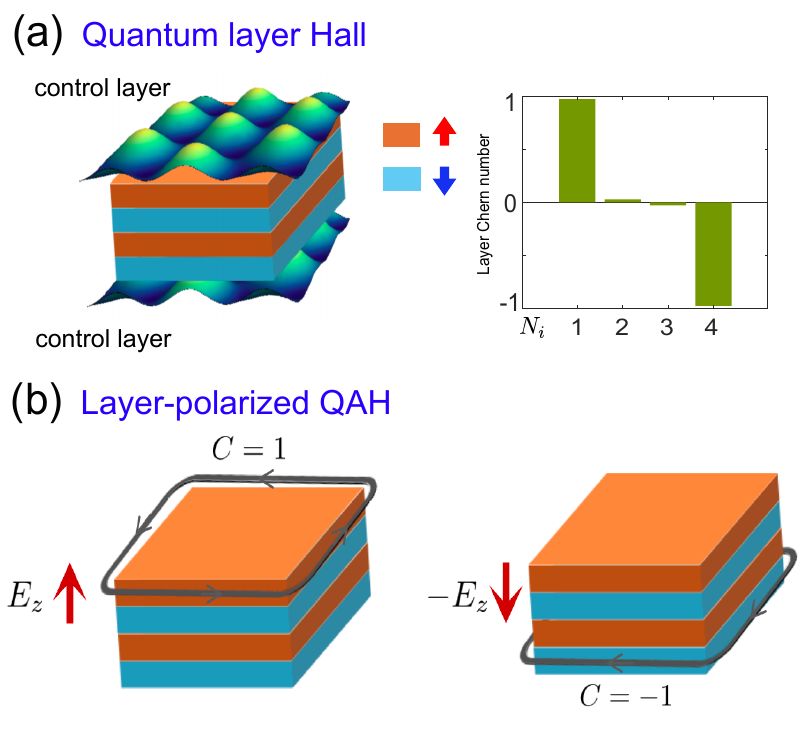}
		\caption{(a) Schematic illustration of the quantum layer Hall effect (QLHE) via moir\'{e} engineering. Dual control layers on both top and bottom surfaces preserve $PT$ symmetry. The layer resolved Chern number is shown to be nearly quantized for the top and bottom layers ($N_1$ and $N_4$). (b) An applied $E_z$ breaks $PT$ and selects which layer hosts chiral edge states.  }
		\label{fig:fig2}
\end{figure}

These developments naturally raise the question of whether the electrically tunable LHE can be quantized into a quantum layer Hall effect (QLHE), in which the opposite layer Hall responses are quantized and locked by opposite layers (see Fig.~\ref{fig:fig1}(b)), thereby providing an experimentally feasible route toward electrically controlled, layer-polarized QAH phases in compensated AFMs. In this work, we demonstrate that moir\'{e} engineering provides a general solution to realize the QLHE in MBT thin films. Moir\'{e} superlattices can be fabricated by an integrated patterned dielectric substrate~\cite{forsythe2018band,li2021anisotropic,barcons2022engineering,wang2024dispersion,tan2024designing,yang2024topological}, which induces a periodic electrostatic potential with a period of tens of nanometers in the target material. Our design treats this dielectric substrate as the control layer as shown in Fig.~\ref{fig:fig2}(a). 

By placing such control layers on an even-layer MBT flake, the resulting periodic potential reconstructs the band topology into flattened moir\'{e} minibands, quantizing the LHE into a QLHE. A perpendicular electric field then selects either the top- or bottom-layer quantized Hall channel, generating a layer-polarized QAH phase as shown in Fig.~\ref{fig:fig2}(b). Importantly, the reduced energy scale of the flattened moir\'{e} minibands substantially lowers the electric field required for layer polarization, providing a practical route toward electrically programmable QAH phases without net spin magnetization. Our work establishes moir\'{e}-engineered topological AFMs as a versatile platform for realizing the QLHE, extending the family of quantum Hall phenomena from spin and valley to the layer degree of freedom and paving the way for a new class of layer-based topological devices~\cite{dai2022quantum,anirban2023quantum,zhai2023time,li2024dissipationless}.

\begin{figure*}
		\centering
		\includegraphics[width=1.0\linewidth]{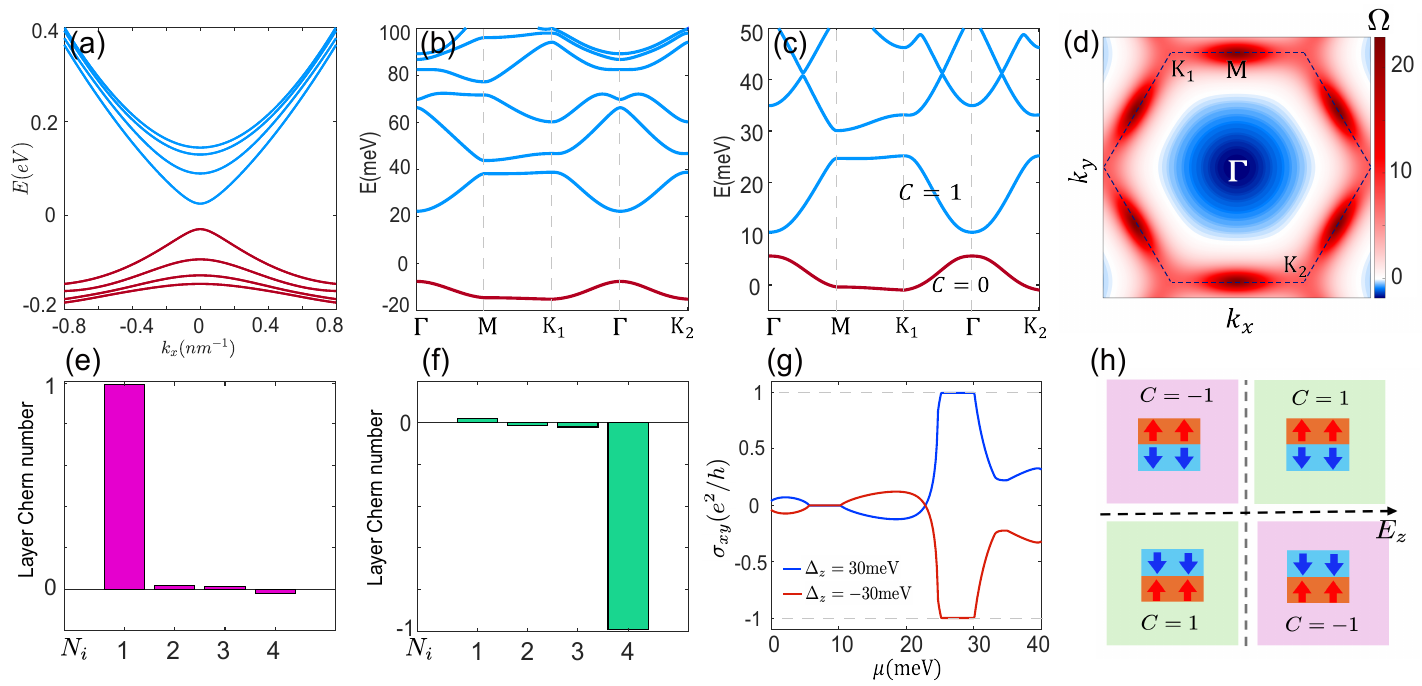}
`		\caption{(a) Band structure of tetralayer MBT thin film. The conduction and valence bands are labeled by blue and red. (b) The moir\'{e} bands are formed by adding control layers to bottom and top surfaces. (c) Applying vertical displacement field by setting $\Delta_z=30$ meV. (d) The momentum-space distribution of Berry curvature for the lowest conduction band with Chern number $C=1$. (e) and (f) The layer-resolved Chern number for $\Delta_z=\pm30$ meV. (g) The anomalous Hall conductivity as a function of Fermi energy. (h) The phase diagram of QLHE with respect to AFM states and $E_z$. Parameters for moir\'{e} potential: $V_0=15$meV, $L_M=30$nm.} 
		\label{fig:fig3}
\end{figure*}

{\emph {Model Hamiltonian and design protocols.}}---As pointed out in Ref.~\cite{wang2015quantized,gao2021layer}, even-layer MBT flakes possess fully compensated AFM order with zero net magnetization, yet they host nontrivial topological properties protected by the combined $PT$ symmetry. We describe the bare Hamiltonian of the $N_l$-layer MBT film as $H_0$ (the details of $H_0$ is present in Supplementary Material~\cite{NoteX}). For the case of a tetralayer MBT film ($N_l = 4$) as an illustrative example, we plot the corresponding band structure in Fig.~\ref{fig:fig2}(a), which exhibits doubly degenerate energy bands due to $PT$ symmetry.

To capture the effects of periodic moir\'{e} potential $U(\bm{r})$ induced by the control layer, we can write down the generic model Hamiltonian  $H=\int d\bm{r}\psi^{\dagger}(\bm{r})\mathcal{H}(\bm{r})\psi(\bm{r})$. Here,
\begin{equation}
	\mathcal{H}(\bm{r})=H_0+U(\bm{r}).\label{moire}
\end{equation}
We then choose a $C_6$-periodic moir\'e potential to illustrate our main results, which is given by 
\begin{equation}
\label{eq:eq_moire}
U(\bm{r})=2V_0\sum_{j=1,3,5}\cos(\bm{G}_j\cdot\bm{r}+\phi),
\end{equation}
with moir\'e reciprocal wave vectors $\bm{G}_{j}=\frac{4\pi}{\sqrt{3}L_{M}}(\cos(\frac{(j-1)\pi}{3}),\sin(\frac{(j-1)\pi}{3}))$. Here $L_M$ is the lattice constant of the moir\'e superlattice . For brevity we set $\phi=0$ throughout this work and the case of $\phi=\pi$ can be obtained by particle-hole transformation. To describe the effect from vertical displacement field, we set $\Delta_z$ as the potential difference between the top and bottom layers.

In our design protocol as depicted in Fig.~\ref{fig:fig2}(a), control layers are placed on both surfaces of the even-layer MBT film while preserving $PT$ symmetry. Under this condition, the moir\'e bands remain doubly degenerate, and the Chern number of each moir\'e band is identically zero. Figure~\ref{fig:fig3}(b) shows the corresponding moir\'e band structure at a moi\'{e} potential strength $V_0 = 15$ meV. When a weak out-of-plane gate electric field $\Delta_z = 30$ meV is applied, the $PT$ symmetry is broken. As shown in Fig.~\ref{fig:fig3}(c), this symmetry breaking lifts the band degeneracy and induces a nontrivial Chern band with Chern number $C = 1$ for the lowest conduction band, while the top valence band remains trivial ($C = 0$). We then plot the momentum-space Berry curvature $\Omega(\bm{k})$ for the Chern band in Fig.~\ref{fig:fig3}(d). When the chemical potential lies where the first moir\'{e} conduction band is filled, QAH in antiferromagnetic MBT film can be realized.

It is important to note that the QAH originates from the hidden layer-locked Berry curvature~\cite{chen2024layer}. At $\Delta_z = 0$, electrons are deflected to opposite sides of the top and bottom layers identically due to their opposite Berry curvatures. A $PT$-breaking field polarizes the electrons onto one surface: the electric field pushes the bottom-surface band to higher energy while lowering the top-surface band, thereby confining the quantized Hall current almost entirely to the top layer. To see this more clearly, we can examine the layer resolved transport current $\bm{J}_i=-e\int_{\bm{k}}(\bm{v}_{\bm{k}}^i-\bm{E}\times \bm{\Omega}_i)f_{\bm{k}}$. The layer-resolved Chern number $C_{i}$ can be calculated by by integrating the layer-resolved Berry curvature $\bm{\Omega}_i$ as
\begin{equation}
  C_{i} = -\frac{1}{\pi}\sum_{n\in occ} \int d^2 \bm{k} \mathrm{Im} \frac{\langle u_{n\mathbf{k}}| v_x^i|u_{m\mathbf{k}}\rangle \langle u_{m\mathbf{k}}|v_y|u_{n\mathbf{k}}\rangle}{(E_{m}-E_{n})^2},
\end{equation}
where $v_{x}^i=\{ v_x,P_i\}/2$ with $\hat{P}_i$ representing the projecting operator onto the $i$-th layer. The summation runs over all occupied bands in the first moir\'{e} Brillouin zone.

To verify the layer polarized nature of the QAH state, we compute the layer-resolved Chern numbers $C_{i}$ for layers $i = 1,2,3,4$ (where $i=1$ denotes the top layer and $i=4$ the bottom layer), as shown in Figs.~\ref{fig:fig3}(e) and (f). The case for $\Delta_z = 0$ is already shown in Fig.~\ref{fig:fig3}(a) with zero total Chern number but quantized layer resolved Chern numer on the top and bottom layers. At $\Delta_z = 30$ meV, we find $C_{N_1} \approx 1$, confirming that the QAH state is fully polarized on the top layer. Reversing the direction of $\Delta_z$ yields $C_{N_4} \approx 1$, implying that the polarization switches to the bottom layer. Owing to weak but finite interlayer coupling, the Chern numbers for the intermediate layers ($i = 2,3$) are not strictly zero; however, their magnitudes remain negligibly small (less than $0.01$) and do not affect the quantized layer transport. We also plot the Hall conductance $\sigma_{xy}$ versus chemical potential $\mu$ in Fig.~\ref{fig:fig3}(g) at $\Delta_z=\pm30$ meV. It can bee seen that a quantized Hall plateau emerges, and $\sigma_{xy}$ changes sign when the polarity of $\Delta_z$ is flipped. Finally, the layer-resolved topological effect also depends on the two degenerate AFM configurations. As shown in Fig.~\ref{fig:fig3}(h), the QAH conductance can be controlled by both the electric field and the AFM states, implying that the QLHE is fundamentally locked to both the layer polarization and N\'{e}el order.

\begin{figure}
		\centering
		\includegraphics[width=1.0\linewidth]{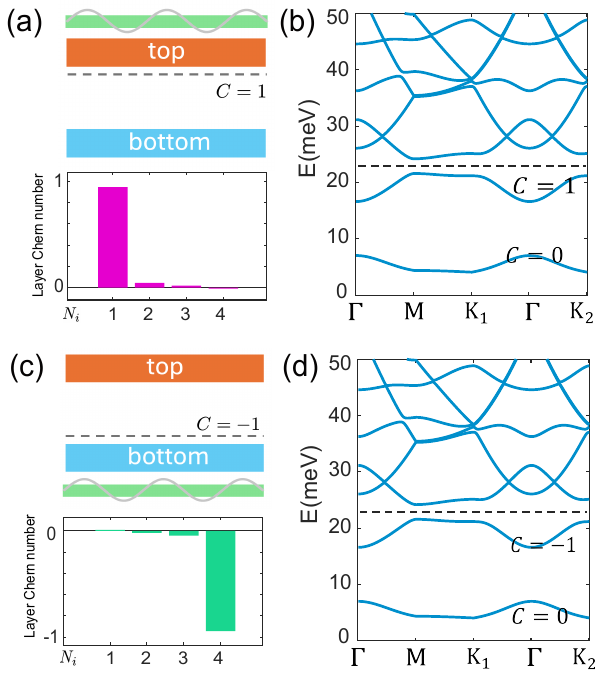}
		\caption{ (a) The control layer is placed on the top surface, resulting in the $C=1$ for the lowest conduction band with the layer-resolved Chern number is shown in the lower panel. (b) The corresponding moir\'{e} band structure. (c) and (d) The control layer is placed on the bottom surface. Parameters for moir\'{e} potential: $V_0=15$meV, $L_M=50$ nm.} 
		\label{fig:fig4}
\end{figure}

We now discuss the accessibility of the layer-polarized QAH without an external electric field in our second protocol. As schematically shown in Fig.~\ref{fig:fig4}(a), placing a control layer on the top surface of the MBT film breaks global $PT$ symmetry, leading to the emergence of topological moir\'{e} minibands [Fig.~\ref{fig:fig4}(b)]. The layer-resolved topological properties can be again revealed by examining the layer-dependent Chern numbers. As shown in the lower panel of Fig.~\ref{fig:fig4}(a), $C_{N_1} \approx 1$, while contributions from other layers are negligible, implying that chiral edge state propagates primarily along the top-layer edge. When the control layer is instead placed on the bottom surface, the band structure [Fig.~\ref{fig:fig4}(d)] is identical to that in Fig.~\ref{fig:fig4}(b), but with $C_{4} \approx 1$, meaning that the chiral edge state localizes on the bottom layer.

{\emph {Emergent layer $U(1)$ gauge field.}}---After numerically demonstrating the formation of the QLHE with layer-polarized Chern bands in moir\'{e} MBT multilayers, we now provide a physical interpretation of the numerical observations above. For an even-layer MBT film, the low-energy physics can be effectively described by the two-surface model~\cite{mei2024electrically}, which reads
\begin{equation}
\label{eq:eq_twosurface}
H=v_0(p_y s_x-p_x s_y)\tau_z+Ms_z\tau_z+\Delta_ts_0\tau_x+U(\bm{r})s_0\tau_0,
\end{equation}
where $M$ denotes the layer-contrasted exchange field and $\Delta_t$ is the interlayer coupling. $v_0$ is the Fermi velocity. Using the projection method~\cite{hu2023berry}, we obtain the effective Hamiltonian for the conduction bands
	\begin{equation}
		\label{Eq2}	H_{eff}=\frac{1}{2m^*}(\bm{p}-e\chi\bm{A})^2+U(\bm{r})+M,
	\end{equation}
where $m^*$ is the effective mass with $m^* = M/v_0^2$, and $\chi = \pm 1$ denotes the top (bottom) layer. In the derivation, we neglect $\Delta_t$ as it is sufficiently weak for a multilayer MBT film. The moir\'{e} potential $U(\bm{r})$ effectively hybridizes the conduction and valence bands. In particular, we find that the scalar moir\'{e}  potential provides a gauge field $\bm{A}(\bm{r})$ that has the opposite sign for the top and bottom layers, yielding
\begin{equation}
\bm{A}(\bm{r})=\frac{\hbar}{4 e M}(\partial_yU(\bm{r}),-\partial_x U(\bm{r})).
\end{equation}
The vector potential $\bm{A}(\bm{r})$ obeys Coulomb gauge $\nabla\cdot \bm{A}(\bm{r})=0$. One can regard $\bm{A}$ as a gauge potential so that we define the pseudomagnetic field $B_{ps}(\bm{r})$ as
	\begin{equation}
		\label{Eq4}
		B_{ps}(\bm{r})=\partial_x A_y-\partial_y A_x=-\chi\frac{\hbar}{4eM}\nabla^2 U(\bm{r}),
	\end{equation}
The strength of the pseudomagnetic field is primarily determined by the energy gap $M$ and the moir\'{e} potential $U(\bm{r})$. Analogous to the Haldane model~\cite{haldane1988model,xie2022valley}, this periodic pseudomagnetic field opens a topological gap with opposite signs on the top and bottom layers. We can see this more clearly by examining the magnetic flux $\Phi_{\rm{ps}}^{\chi}=\int_{\rm{cell}}B_{ps}^{\chi}d^2\bm{r}$, which can be further written as
\begin{equation}
\Phi_{\rm{ps}}^{\chi}=-\chi\frac{\hbar}{4eM}\oint_{\partial\rm{cell}}\bm{\nabla}U\cdot\bm{n}dl=0.
\end{equation}

\begin{figure}
		\centering
		\includegraphics[width=1.0\linewidth]{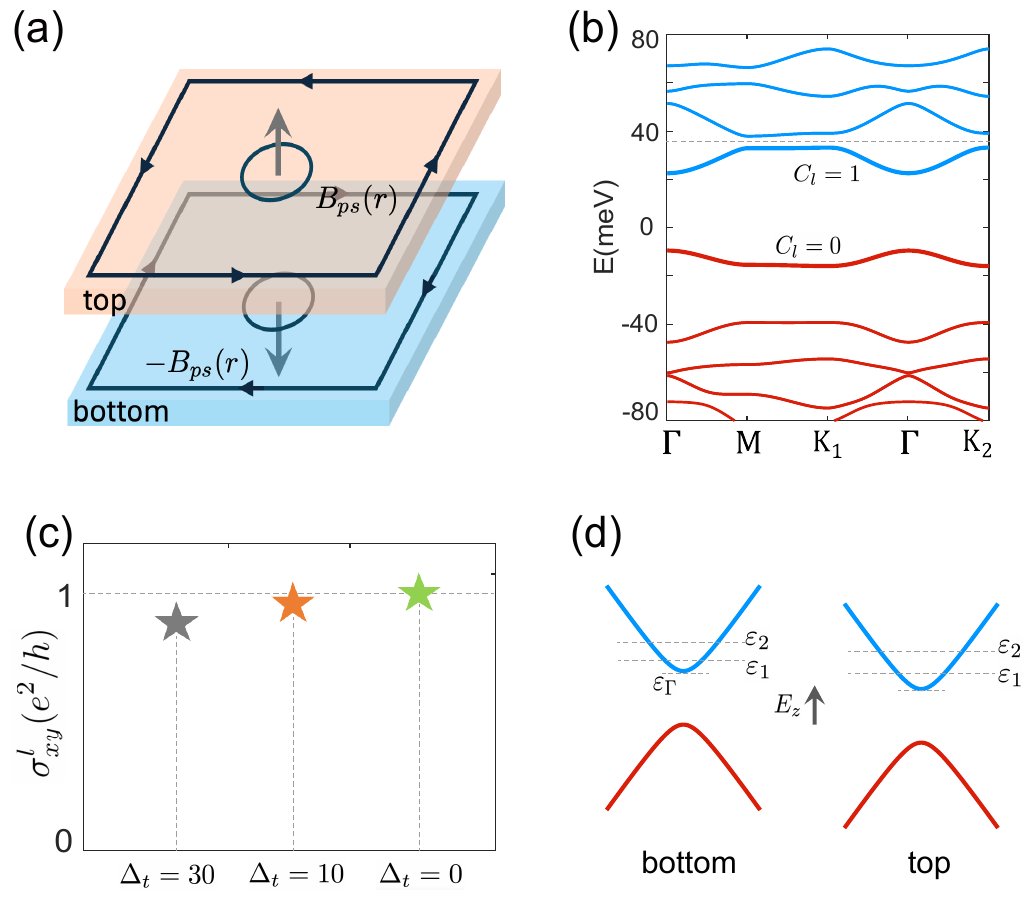}
		\caption{ (a) Schematic illustration of the layer-contrasted pseudomagnetic field. Two chiral edge states propagate in opposite directions relative to each other. (b) Moir\'{e} minibands for $V_0 = 15$ meV and $L_M = 30$ nm. (c) Layer Hall conductance $\sigma_{xy}^l$ with various interlayer coupling strength $\Delta_t$. (d) Schematic of the electric-field-driven topological phase transition. Other parameters: $\hbar v_0 = 0.3$ eV$\cdot$nm and $M = 30$ meV.} 
		\label{fig:fig5}
\end{figure}

The proposed QLHE shares similarities with the quantum spin Hall in the nature of their respective $U(1)$ symmetries. The QLHE, at first glance, appears to have an exact $U(1)$ layer-pseudospin symmetry $[H, \tau_z] = 0$ when $\Delta_t = 0$ due to the spatial separation of top and bottom surface states. One can define a layer Chern number $C_l = (C_t - C_b)/2$. The lowest conduction band carries $C_l = 1$, whereas the highest valence band has $C_l = 0$ (see Fig.~\ref{fig:fig4}(b)). Finite interlayer coupling $\Delta_t$ causes $\sigma_{xy}^l$ to deviate from quantized value, as shown in Fig.~\ref{fig:fig5}(c). We find that even for a large $\Delta_t = 30$ meV, $\sigma_{xy}^l \approx 0.9e^2/h$ remains close to the quantized value. Nevertheless, interlayer coupling in MBT decreases exponentially with film thickness, offering a straightforward experimental knob to suppress such a symmetry breaking term. In this moir\'{e}-engineered regime, the layer Chern number $C_l$ becomes well-defined and quantized, leading to the moir\'e QLHE with a quantized layer Hall conductance $\sigma_{xy}^{l} = e^2/h$.

To understand the electric field driven topological phase transition, we performed the perturbation approach to examine the energy levels near the Brillouin zone corners. To achieve the layer-polarized QAH phase, we need the energy difference between the top and bottom surfaces to overcome the bandwidth, yielding 
\begin{equation}
\Delta_z \gtrsim \varepsilon_0-V_0
\end{equation}
Here $\varepsilon_0=\hbar^2|K|^2/2m^* \sim 1/L_M^2$. Thus, moir\'{e} engineering substantially reduces the critical gate electric field required to achieve the QAH phase, allowing a weak electric field to induce the Chern insulating state in a MBT thin film. For the second protocol that the moir\'{e} control layer is placed on one surface, we need the moir\'{e} potential to be large enough to overcome the bandwidth $\varepsilon_0$ such that $U_0>\varepsilon_0$. This is because the energy level of $\varepsilon_\Gamma$ is shifted to  $\varepsilon_\Gamma-V_0$.

\begin{figure}
		\centering
		\includegraphics[width=1.0\linewidth]{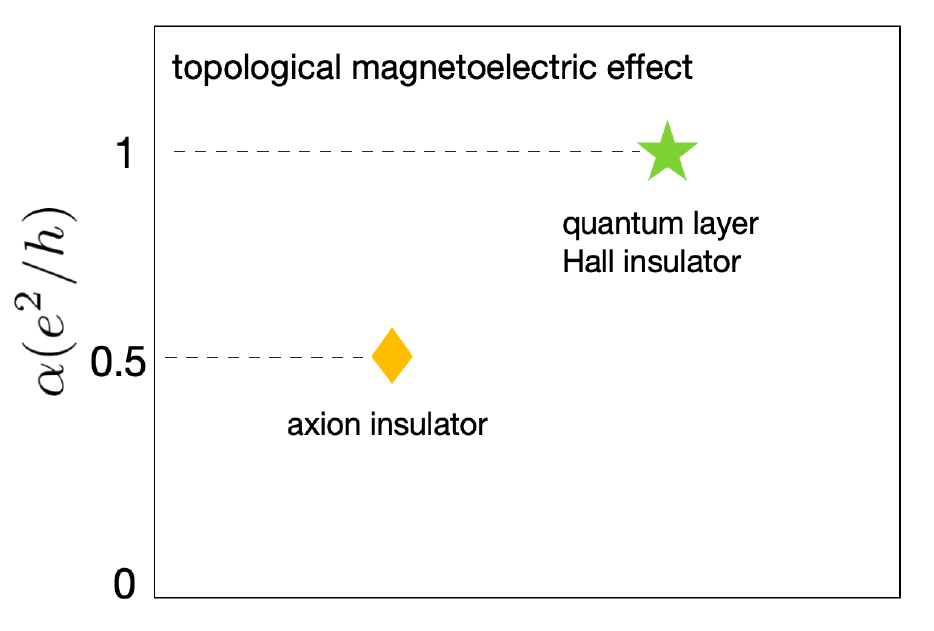}
		\caption{Topological magnetoelectric effect in axion insulator and quantum layer Hall insulator. $\alpha$ denotes the magnetoelectric coefficient.} 
		\label{fig:fig6}
\end{figure}

{\emph {Topological magnetoelectric response.}}---Up to now, we have established that the layer polarized QAH is the transport manifestation of the QLHE. In $PT$-symmetric moir\'{e} MBT flake with dual control layers (Fig.~\ref{fig:fig2}(a)), a quantum layer Hall insulator emerges when the first moir\'{e} conduction band is fully filled. It serves as the parent state of the layer-polarized QAH in collinear topological AFMs. However, since layer resolved transport measurement techniques are not yet available, and even multilayer MnBi$_2$Te$_4$ is only nanometers thick, the layer Chern number cannot be directly read out from transport measurement. Is there an alternative way to identify the quantum layer Hall insulator? Here, we find that the topological magnetoelectric effect provides a definitive solution. Notably, the $PT$-symmetric moir\'{e}-dressed MBT thin film exhibits a topological magnetoelectric response  different from that of the pristine one. The magnetoelectric coefficient $\alpha$ is given by~\cite{essin2009magnetoelectric,mei2024electrically}
\begin{equation}
\alpha=\frac{e^2}{h}\sum_{n\in \mathrm{occ}} C_l(n), 
\end{equation}
where $C_l(n)$ is the layer Chern number for a given band $n$. $\alpha$ describes the out-of-plane electric field induced orbital magnetization via $M_{z}=\alpha E_z$. In an axion insulator, $\alpha$ is half quantized to $e^2/(2h)$. This arises from the layer-resolved Chern numbers $C_{\text{top}} = 1/2$ and $C_{\text{bottom}} = -1/2$, which yield a zero net Hall conductance but a finite axion angle $\theta = \pi$~\cite{essin2009magnetoelectric,wang2015quantized,qiu2025observation,hu2026orbital,mei2024electrically}. In contrast, the QLHE proposed here exhibits a quantized response with $\alpha=e^2/h$, reflecting layer Chern numbers $C_{\text{top}} = 1$ and $C_{\text{bottom}} = -1$. This doubling of the layer Chern number from $\pm 1/2$ to $\pm 1$ is the direct consequence of the moir\'{e}-induced QLHE. Consequently, the QLHE not only enables a quantized layer Hall conductance $\sigma_{xy}^{\text{l}} = e^2/h$ with zero net charge Hall response but also doubles the magnetoelectric coupling, providing a distinct experimental signature (see Fig.~\ref{fig:fig6}).

{\emph{Conclusion and discussion.}}---In this work, we have presented a general theoretical framework for engineering the QLHE in topological AFM thin films. Our findings point to several promising directions. First, the concept of a moir\'{e}-induced layer-locked pseudomagnetic field can be extended to other antiferromagnetic topological insulators, particularly the broader family of MnBi$_2$Te$_4$-type materials~\cite{he2020mnbi2te4,han2025layer}. Second, the designed moir\'{e} potential can be generalized to other lattice patterns, such as the Kagome geometry~\cite{wang2024dispersion} (see Supplementary Material for further details). Third, the interplay between moir\'{e} engineering and electron correlations in the resulting flat minibands may give rise to fractional layer Hall states, analogous to the fractional quantum anomalous Hall effect recently observed in moir\'{e} transition metal dichalcogenide heterobilayers~\cite{cai2023signatures,xu2023observation,zeng2023thermodynamic,park2023observation,kang2024evidence,lu2024fractional}. Finally, the vanishing stray field inherent to the QLHE makes it particularly attractive for engineering superconducting devices and exploring proximity-induced transport phenomena, such as the Josephson effect~\cite{sun2024anomalous,zhang2025observation}. We anticipate that our work will stimulate experimental efforts to realize the moir\'{e} QLHE and inspire further explorations of layer-based topological phenomena in moir\'{e}-engineered antiferromagnetic systems.


%

\end{document}